\documentclass[prl,twocolumn,amssymb,amsmath,aps,floatfix]{revtex4}
\newcommand{\be}{\begin{equation}}
\newcommand{\ee}{\end{equation}}
\newcommand{\bea}{\begin{eqnarray}}
\newcommand{\eea}{\end{eqnarray}}

\usepackage{graphicx}
\usepackage{epsfig}
\usepackage{bbold}
\usepackage{mathtools}

\begin{document}

\title{Fractional quantum Hall states of photons in an array of dissipative coupled cavities}
\author{R. O. Umucal\i lar}
\email{onur@science.unitn.it}
\affiliation{INO-CNR BEC Center and Dipartimento di Fisica, Universit\`a di Trento, I-38123 Povo, Italy}
\author{I. Carusotto}
\email{carusott@science.unitn.it}
\affiliation{INO-CNR BEC Center and Dipartimento di Fisica, Universit\`a di Trento, I-38123 Povo, Italy}

\date{\today}

\begin{abstract}
We report a theoretical study of the collective optical response of a two-dimensional array of nonlinear cavities in the impenetrable photon regime under a strong artificial magnetic field.
Taking advantage of the non-equilibrium nature of the photon gas, we propose an experimentally viable all-optical scheme to generate and detect strongly correlated photon states which are optical analogs of the Laughlin states of fractional quantum Hall physics.
\end{abstract}

\maketitle

In the latest years, the hydrodynamic properties of quantum fluids of light have attracted a strong interest from both theoretical and experimental points of view, with the demonstration of superfluid flow in degenerate quantum gases of dressed photons (the so-called exciton-polaritons) in planar semiconductor microcavities~\cite{Pol_Superfl_Exp} and the subsequent observation of vortices and solitons in the wake of a strong defect~\cite{Pol_Vort_Exp}. All these experiments were performed in a dilute gas regime where a mean-field description based on a generalized Gross-Pitaevskii equation is accurate~\cite{BECbook,GPE}.

Going beyond this regime requires that the underlying medium show a sufficiently large optical nonlinearity to induce strong effective interactions between the photons. A first step in this direction has been the observation of strong antibunching in light emission from single-mode cavities in both the visible~\cite{antibunch-visible,photonic_crystal} and microwave domains~\cite{antibunch-CQED} via the so-called photon blockade effect.
The present experimental challenge is to scale up the impenetrable photon regime to arrays of many coupled cavities, for which theoretical work has anticipated the onset of different kinds of strongly correlated photon states, from Mott insulators~\cite{Hartmann}, to Tonks-Girardeau gases in one-dimensional geometries~\cite{TGfiber,TGlatt}.

In the meanwhile, several proposals have appeared for generating artificial magnetic fields for neutral quantum particles.
The Berry phase~\cite{Berry} accumulated by an optically dressed atom while adiabatically moving in space can be described in terms of an artificial gauge field~\cite{Dalibard}; the nucleation of quantized vortices in a dilute Bose-Einstein condensate under the effect of such a field was observed in~\cite{Spielman}. Many authors have speculated on the possibility of observing quantum Hall liquid states in strongly  interacting atomic gases in free space~\cite{FQHfree} or optical lattices~\cite{Lukin,Palmer,Bhat}.
Very recently, the extension of this idea to photons has been theoretically investigated in a number of configurations, for instance, arrays of coupled optical cavities confining single atoms~\cite{Bose}, microwaves in circuit-QED devices~\cite{Girvin}, solid-state photonic devices in the infrared or visible spectral range~\cite{Hafezi,Keeling,Umu}.

The present paper reports a theoretical study of the optical response of a coupled cavity system to a coherent laser field in a regime where impenetrable photons experience a strong artificial magnetic field.
Our theoretical description is based on a generic model that can be used to describe a number of different physical systems, ranging from macroscopic optical cavities containing atoms~\cite{antibunch-visible}, to photonic crystal cavities~\cite{photonic_crystal} and circuit QED devices~\cite{antibunch-CQED}.
In contrast to prior work~\cite{Bose}, we take full advantage of the driven-dissipative nature of the photon gas~\cite{footnote_Dissipation} to propose an all-optical protocol to generate and characterize strongly correlated photon states which are analogs of the fractional quantum Hall (FQH) states of electrons in two-dimensional geometries under a strong magnetic field~\cite{FQH_review}. The quantum Hall nature of the generated state is assessed in terms of its overlap with the Laughlin wave function~\cite{Laughlin_wf}. We anticipate that detailed information on the microscopic structure of the many-body wave function can be experimentally extracted from the field quadratures of the secondary emission from the device.

We consider a two-dimensional square lattice of coupled cavities under a uniform artificial magnetic field. Each cavity is assumed to sustain a single photonic mode, to be coupled to its nearest neighbor by photon tunneling, and to exhibit a large optical nonlinearity leading to strong on-site interactions between photons. The isolated system can then be described by the standard single-band Bose-Hubbard Hamitonian~\cite{TGlatt,Umu}:
\be
\label{Magnetic_Hamiltonian}
H_0 = \sum_i \hbar \omega_\circ \hat{b}^\dag_i \hat{b}_i -\hbar J\sum_{\langle i,j\rangle} \hat{b}_i^{\dag}\hat{b}_j e^{i\varphi_{ij}}+\hbar\frac{U}{2}\sum_i \hat{n}_i(\hat{n}_i-1),
\ee
where $\hat{b}_i^{\dag}$ ($\hat{b}_i$) is the bosonic creation (annihilation) operator for site $i$ and  $\hat{n}_i = \hat{b}_i^{\dag}\hat{b}_i$ is the corresponding number operator. The effect of the artificial magnetic field is included via the tunneling phase $\varphi_{ij}$; $J$ is the tunneling strength between nearest neighbor sites $\langle i,j\rangle$, $U$ is the on-site interaction energy, and $\omega_\circ$ is the natural cavity frequency.

In the non-interacting case $U/J=0$, the spectrum of the isolated system Hamiltonian $H_0$ for an infinite lattice is the Hofstadter butterfly \cite{Hofstadter} which appears as a fractal structure in the energy versus $\alpha=(2\pi)^{-1}\, \mathrlap{\,\circlearrowleft}\sum \varphi_{ij}$ plane, where the loop sum is performed around a plaquette of the lattice. For a real magnetic field $B$ and a lattice spacing $a$, $\alpha$ would correspond to the number of magnetic flux quanta per plaquette, $\alpha = Ba^2/(h/e)$.
In the presence of finite on-site repulsive interactions, for filling $\nu=1/2$, the bosonic Laughlin wave function turns out to be the almost exact ground-state wave function in the continuum limit $\alpha \ll 1$ \cite{FQHfree,Lukin,Palmer} and to capture the essential features of the ground state up to a critical value of $\alpha\sim0.3$ for sufficiently strong interactions $U/J\gtrsim 5$ ~\cite{Lukin}.

To accurately describe a photonic system, one has to account for the finite photon lifetime and for the optical pump that is used to continuously replenish the photon gas~\cite{TGlatt}. A coherent pump can be included in the model as an additional Hamiltonian term in the form $H_{\rm drive}=\sum_i \big[\hbar F_i(t)\, \hat{b}_i^{\dag}+\textrm{h.c.}\big]$. Here we restrict our attention to the simplest case of a monochromatic pump at frequency $\omega_p$ that drives all sites with the same amplitude, $F_i(t) = \bar{F}\,e^{-i\omega_p t}$.
Photon losses at a rate $\gamma$ are described at the level of the master equation $d\rho/dt = -\frac{i}{\hbar}[H_0+H_{\rm drive},\rho]+\mathcal{L}[\rho]$ with a dissipative term in the Lindblad form $\mathcal{L}[\rho]=\gamma \sum_i \big[\hat{b}_i\rho \hat{b}_i^{\dag}-(\hat{b}_i^{\dag}\hat{b}_i\rho+\rho \hat{b}_i^{\dag}\hat{b}_i)/2\big]$~\cite{QuantumOptics}. As in previous work~\cite{TGlatt} the steady-state density matrix $\rho_{ss}$ is obtained from a numerical determination of the stationary point of the master equation $d\rho/dt=0$. More details are given in the Supplemental Material.

We now present the results of our numerical calculations for a $4\times4$ lattice using a basis of states with total photon number $N=0,1,2$ in the hardcore limit $U/J=\infty$, where double occupation of a single site is not allowed. Excitation to higher $N$ states is negligible as long as the pump amplitude is well below saturation $\bar{F}\ll \gamma$. We work in the Landau gauge $\mathbf{A} = (-By,0)$ with an effective magnetic field strength $B$ such that $\alpha = 1/4$. This choice corresponds to $N_\alpha=4$ flux quanta through the whole lattice and therefore to a filling fraction $\nu = N/N_{\alpha} = 1/2$ for an $N=2$ state. Periodic boundary conditions are assumed~\cite{footnote_BCs}.

\begin{figure}[htbp]
\includegraphics[width = \columnwidth]{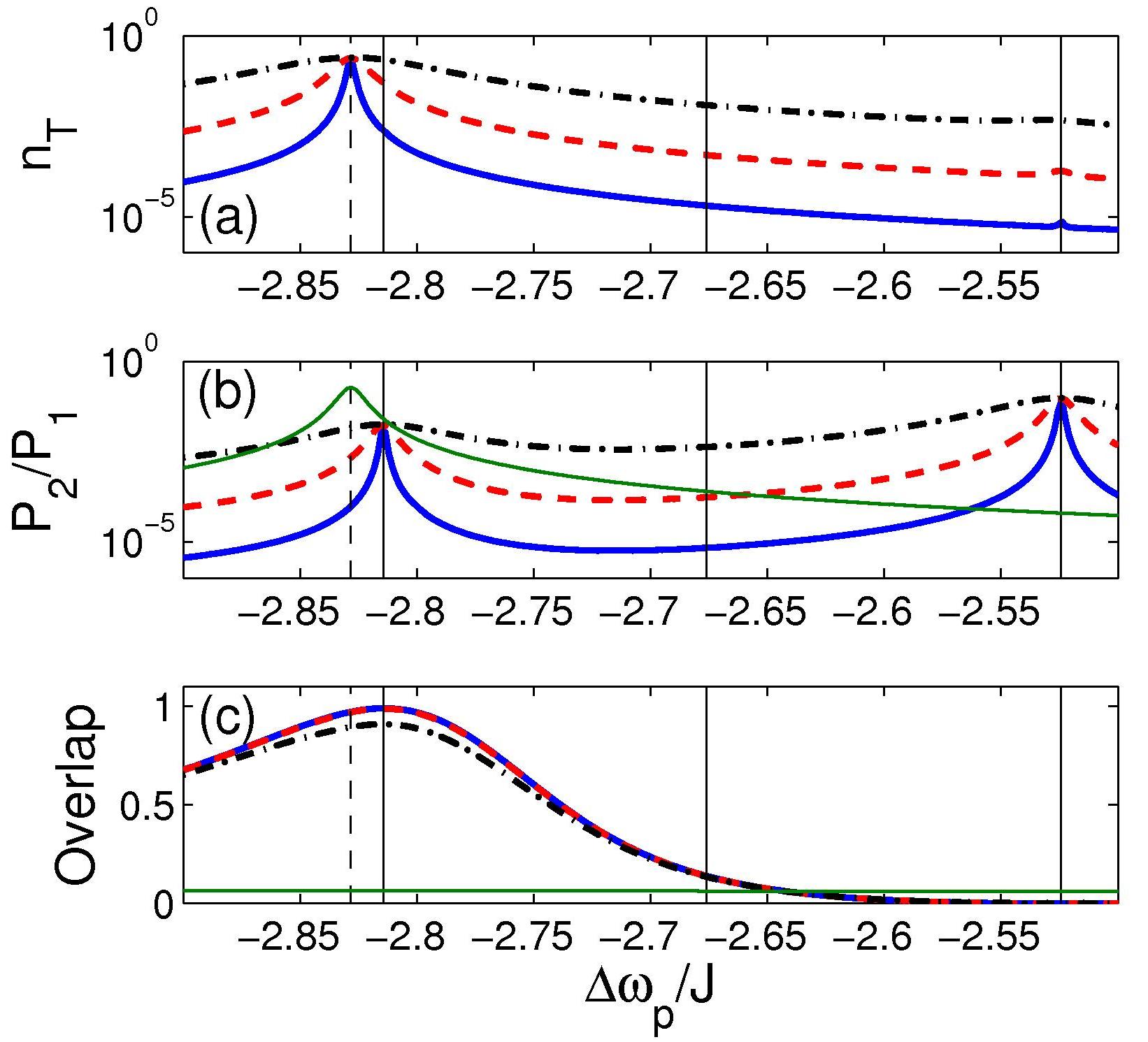}
\caption{(Color online) (a) Expectation value of the total number of particles. (b) Ratio $P_2/P_1$ of the probability of having two particles to one particle in the steady-state.
(c) Overlap $\mathcal{O}$ with the optimized Laughlin wave function.
All curves in (a-c) are plotted as a function of pump frequency $\Delta\omega_p$ for different values of the loss rate $\gamma/J = 0.002$ (blue solid), $0.01$ (red dashed), $0.05$ (black dash-dotted). Pump amplitude is $\bar{F}/\gamma=0.1$ for all cases. A $4\times4$ lattice with periodic boundary conditions is considered in the impenetrable photon limit $U/J=\infty$. Vertical solid (dashed) lines indicate the position of two- (one-) photon transitions as predicted by the eigenfrequencies of $H_0$. The green thin curves in (b) and (c) refer to the non-interacting photon case  $U/J=0$ for $\gamma/J=0.01$.
\label{4x4transmission_overlap}}
\end{figure}

The first observable we consider is the total number of photons $n_T = \sum_i \langle \hat{b}_i^\dagger \hat{b}_i \rangle$ present in the steady state of the system, a quantity which in many geometries is proportional to the total transmitted intensity. This quantity is plotted in Fig.~\ref{4x4transmission_overlap}(a) as a function of the relative pump frequency $\Delta\omega_p = \omega_p-\omega_\circ$ for three different values of the loss rate $\gamma$ and a fixed pump amplitude $\bar{F}/\gamma=0.1$.
The main feature is a strong peak at $\Delta\omega_p/J\approx -2.83$ corresponding to a one-photon transition from the vacuum to the lowest one-particle eigenstate of $H_0$ at energy $\hbar (\omega^{(1)}+\omega_\circ)$: the position of the resonance is at $\Delta\omega_p = \omega^{(1)}$. In addition, on the spectrum for the lowest value of $\gamma$ we can notice a much weaker peak at $\Delta\omega_p/J\approx -2.52$ corresponding to a two-photon transition from vacuum to a two-particle eigenstate of $H_0$ of energy $\hbar (\omega^{(2)}+2\omega_\circ)$: in this case, the position of the resonance is given by $\Delta\omega_p=\omega^{(2)}/2$~\cite{TGlatt}.

In order to isolate the two-photon peaks from the much stronger one-photon background, in Fig.~\ref{4x4transmission_overlap}(b) we plotted the ratio $P_2/P_1$ where the steady-state probability of having $N=1,2$ particles in the system is defined as $P_{1,2}=\textrm{Tr}[\rho_{ss}\Pi_{1,2}]$ in terms of the projectors $\Pi_{1,2}$ onto the one- or two-particle subspaces: in this way, clear peaks appear at half the frequency of the lowest and the third lowest $N=2$ eigenstates of $H_0$. The second lowest $N=2$ eigenstate of $H_0$ does not contribute to this spectrum as the matrix element for the corresponding two-photon transition appears to be very small for the chosen spatial profile of the pump.

A common procedure in the theory of the quantum Hall effect to analyze the physical nature of a state is to calculate its overlap with an ansatz wave function, for example, a Laughlin wave function.
Here we extend this approach to a driven-dissipative system: under a weak pump condition $\bar{F}\ll \gamma$, the component of the density matrix on the $N=2$ subspace $\rho^{(2)} =\Pi_2 \rho_{ss} \Pi_2$ involves a single pure two-particle state, whose wave function is proportional to the two-photon amplitude $\psi_{ij}=\textrm{Tr}[\rho_{ss}\hat{b}_i\hat{b}_j]$, which then plays the role of a two-photon wave function.
The quantum Hall nature of the two-photon peaks can then be assessed by calculating the overlap of $\psi_{ij}$ with the pair of generalized Laughlin wave functions $\Psi^{(1,2)}(z_i,z_j)$ for our toroidal geometry~\cite{Haldane,Read}, the explicit form of which is given in the Supplemental Material. The optimized Laughlin wave function is the linear combination of $\Psi^{(1,2)}$ that gives the maximum value $\mathcal{O}$ of the overlap
\begin{equation}
\mathcal{O}=\sum_{l=1,2}\left[ \frac{\sum_{i,j} |\psi^\ast_{ij} \Psi^{(l)}(z_i,z_j)|^2}{\sum_{i,j} |\psi_{ij}|^2}\right].
\end{equation}

A plot of $\mathcal{O}$ as a function of $\Delta\omega_p$ is shown in Fig.~\ref{4x4transmission_overlap}(c). For the weakest loss rate $\gamma/J=0.002$, the overlap is broadly peaked around the lowest two-photon resonance, with a peak value as high as $98.9\%$  (as usual, perfect overlap would correspond to $\mathcal{O}=100\%$) and then decays to zero as the higher two-photon resonance is approached: this is a clear signature that the lowest two-photon peak indeed corresponds to the transition to a two-particle Laughlin state, while the second peak corresponds to some orthogonal excited state.
For a more realistic loss rate $\gamma/J = 0.05$, the maximum overlap is somewhat reduced by the reduced selectivity of the coherent drive and the consequent mixing with other states; however, its value is still as high as $90.9\%$.
The situation is of course completely different in the $U/J=0$ case of non-interacting photons. In this case, only the one-particle peak at $\Delta\omega_p/J\approx -2.83$ remains visible in the spectrum [thin curve in Fig.~\ref{4x4transmission_overlap}(b)] and the overlap with the Laughlin wave function is dramatically reduced at all frequencies [thin curve in Fig.~\ref{4x4transmission_overlap}(c)].

\begin{figure}[htbp]
\includegraphics[width = 0.54 \columnwidth]{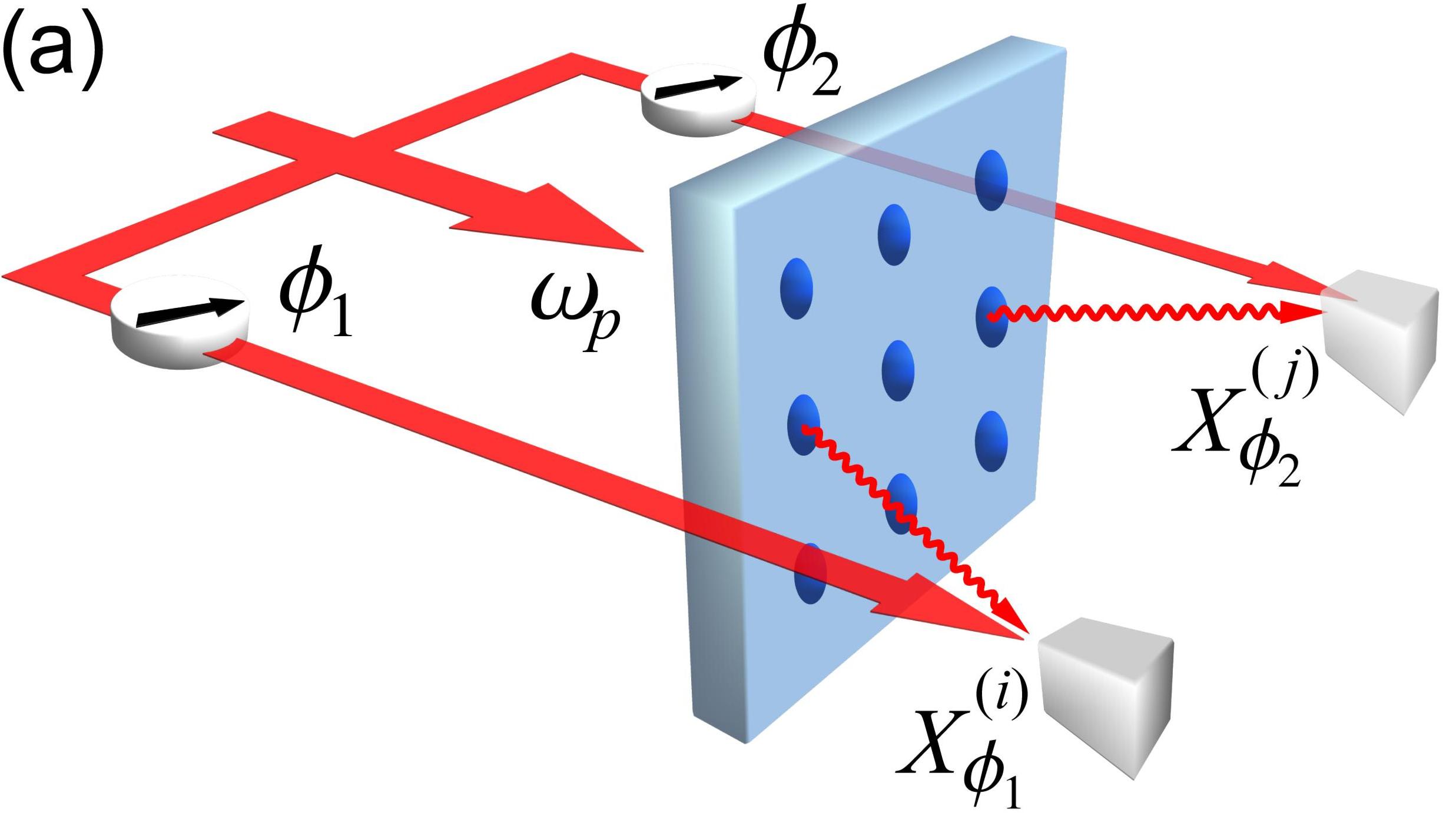}
\includegraphics[width = 0.30 \columnwidth]{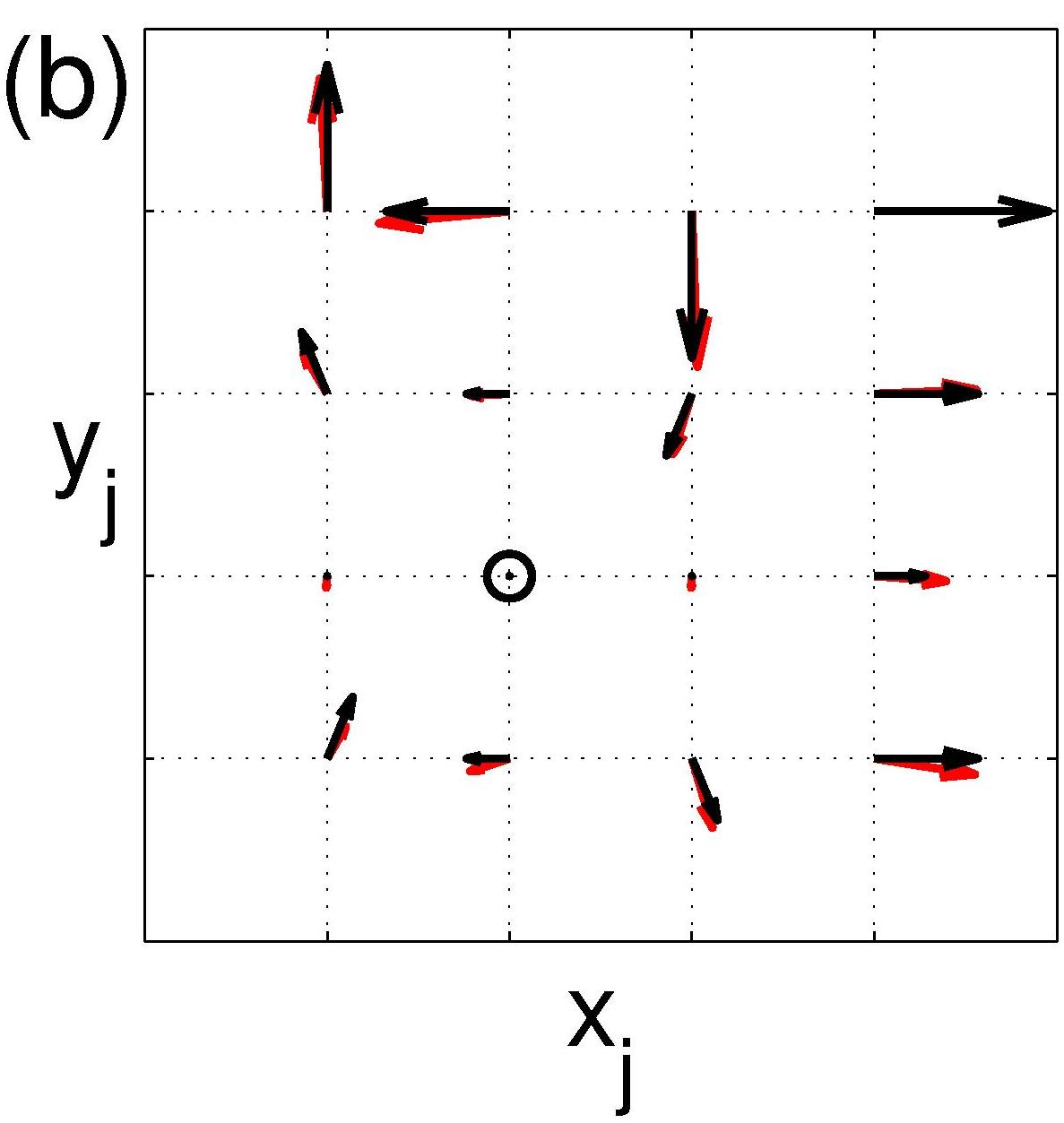}
\caption{(Color online) (a) Sketch of the proposed experimental setup to measure the two-photon amplitude $\psi_{ij}$. (b) $i=i_\circ$ cut of the (normalized) two-photon amplitude $\psi_{i_\circ j}$ (red) and of the optimized Laughlin wave function $\Psi(z_{i_\circ},z_j)$ (black). The two functions are almost indistinguishable. System parameters: $\Delta \omega_p/J=-2.8144$; $\gamma/J=0.03$; $\bar{F}/\gamma=0.1$. $x,y$ axes refer to the $z_j=x_j+iy_j$ coordinate. Reference site $i_\circ$ is marked by a circle. For each lattice site, the arrow indicates the complex amplitude of the wave function.
\label{setup}}
\end{figure}

In contrast to standard condensed matter systems, quantum optical techniques provide experimental access to the overlap $\mathcal{O}$ with the Laughlin state~\cite{footnote_G2}. In the infrared/visible range, the two-photon amplitude $\psi_{ij}$ can in fact be measured with standard homodyne detection techniques~\cite{QuantumOptics} by homodyning the secondary emission from sites $i$ and $j$ with the coherent pump at $\omega_p$ as sketched in Fig.~\ref{setup}. To obtain the different quadratures $X^{(i)}_{\phi_i} = \big(\hat{b}_i e^{-i\phi_i}+\hat{b}_i^\dag e^{i\phi_i} \big)/2$ of the field emitted by site $i$, one has simply to adjust the phase delay $\phi_i$ of the homodyne beam. As quadrature operators for two different sites ($i \neq j$) commute, they can be simultaneously measured and correlated. The full two-photon amplitude $\psi_{ij}$ is finally reconstructed by combining four separate measurements:
\begin{multline}
\langle\hat{b}_i\hat{b}_j\rangle = \langle X^{(i)}_{0}X^{(j)}_{0}\rangle - \langle X^{(i)}_{\pi/2}X^{(j)}_{\pi/2}\rangle \\
+ i \langle X^{(i)}_{0}X^{(j)}_{\pi/2}\rangle + i \langle X^{(i)}_{\pi/2}X^{(j)}_{0}\rangle.
\end{multline}
In the microwave domain of circuit-QED systems, the situation is even simpler as field quadratures can be directly measured using linear amplifiers~\cite{homodyne-CQED}.
A measurement of the two-photon amplitude $\psi_{ij}$ provides an intuitive picture of the complexity of two-photon FQH states. A numerically simulated image of this quantity (actually, a cut of $\psi_{ij}$ for a fixed $i=i_\circ$) is illustrated in Fig.~\ref{setup}(b) and compared with the Laughlin wave function; the agreement is excellent as expected from the high values of the overlap mentioned previously.

\begin{figure}[htbp]
\includegraphics[width = \columnwidth]{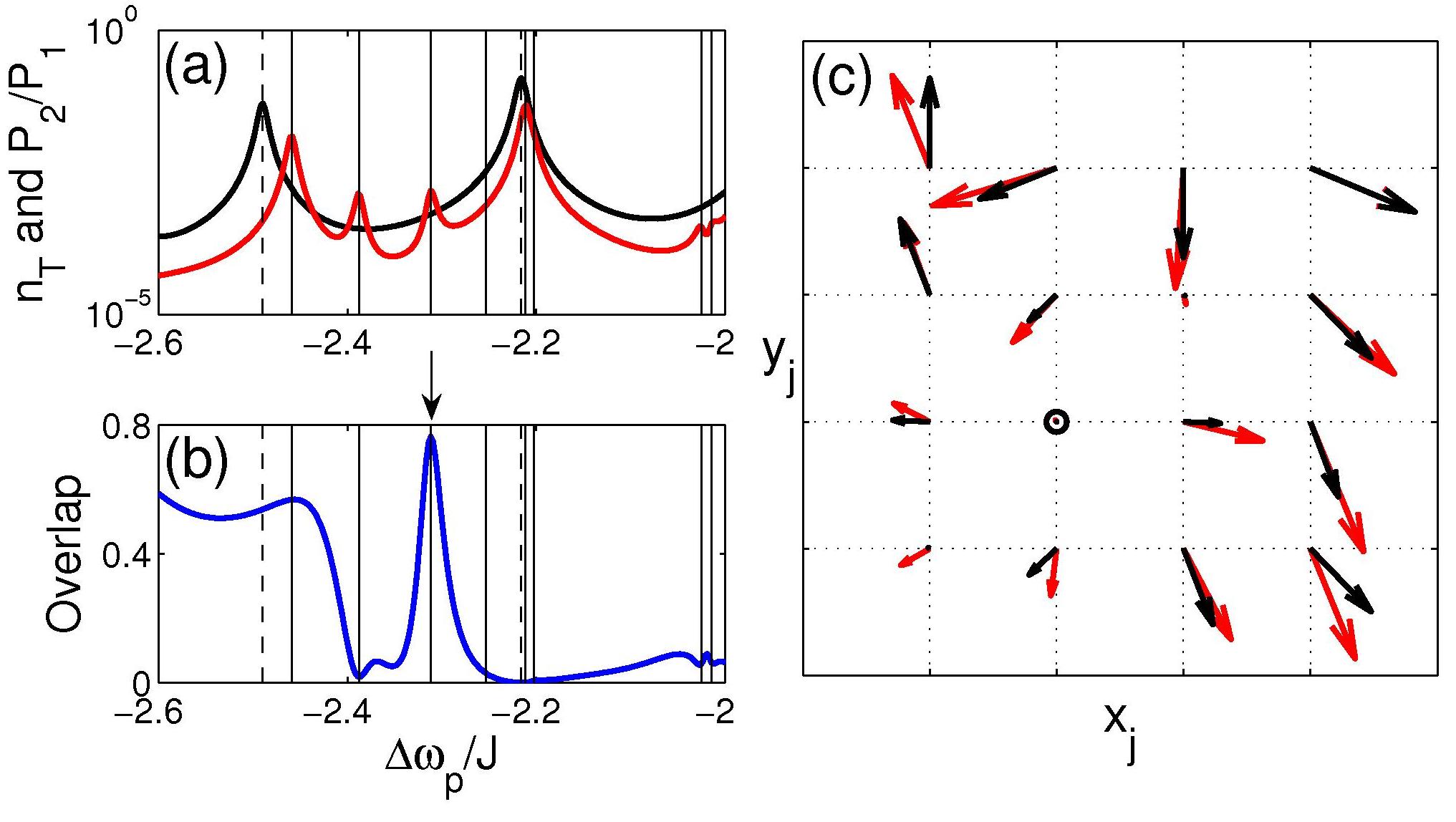}
\caption{(Color online) System with hard-wall boundary conditions and finite interaction strength $U/J = 20$. (a) Expectation value of the total number of particles (black solid curve) and $P_2/P_1$ ratio (red solid curve) as a function of pump frequency. (b) Overlap $\mathcal{O}$ of the two-photon amplitude $\psi_{ij}$ with the Laughlin wave function. Vertical solid (dashed) lines correspond to two- (one-) photon transitions. (c)  $i=i_\circ$ cut of the two-photon amplitude $\psi_{i_\circ j}$ (red) and of the optimized Laughlin wave function (black) for a pump on resonance with the two-photon transition at $\Delta \omega_p/J=-2.3120$ indicated by an arrow in (b); $\gamma/J=0.01$; $\bar{F}/\gamma=0.1$.
\label{4x4transmission_overlap_hard}}
\end{figure}

We conclude the paper by discussing how all the previously described phenomena are robust against expected experimental imperfections and are not restricted to small lattices.
Although periodic boundary conditions may be realized in a photonic system (for instance, by suitably connecting sites on opposite sides with optical fibers), an experimentally much less demanding option is to use hard-wall boundaries. Furthermore, there might be structural defects leading to a small inhomogeneity in the lattice potential and interactions between the particles might be not strong enough to be in the hard-core limit $U/J=\infty$.
Transmission spectra for a more realistic case of $4\times4$ lattice with finite $U/J= 20$ and an energy offset on the order of $\gamma$ on a few lattice sites are shown in Fig.~\ref{4x4transmission_overlap_hard} together with the corresponding overlap with the optimized Laughlin wave function for periodic boundary conditions. In spite of the different hard-wall geometry and the disturbing effect of disorder and finite interactions, the overlap on resonance with the $\Delta \omega_p/J = -2.3120$  two-photon transition is as high as $\sim76\%$. This fact is intuitively visible in the two-photon amplitude plotted in Fig.~\ref{4x4transmission_overlap_hard}(c). Although there are some deviations, the complex structure of the Laughlin wave function is still clearly recognizable.

Calculations for larger systems and more particles can be performed with a more sophisticated Monte-Carlo wave-function technique~\cite{Monte-Carlo} which is detailed in the Supplemental Material and is much more time-consuming. For this reason, we did not calculate the whole spectrum, but we restricted ourselves to a single value of the pump frequency, chosen to be on exact three-photon resonance with the lowest three-particle eigenstate of an isolated $5\times 5$ lattice with a gauge field intensity $\alpha=0.24$ so as to give again $N/N_\alpha=1/2$. For $\gamma/J = 0.01$, $\bar{F}/\gamma = 0.1$, averaging over 500 independent Monte Carlo realizations yields an overlap of $\sim 93\%$ between the three-photon amplitude $\psi_{ijk}=\textrm{Tr}[\rho_{ss}\hat{b}_i\hat{b}_j\hat{b}_k]$ and the Laughlin wave function, meaning that our all-optical technique is able to generate also more complex Laughlin states of more than two particles. Of course $\psi_{ijk}$ can again be measured with the same homodyne techniques by combining eight measurements of the form $\langle X^{(i)}_{\phi_i}X^{(j)}_{\phi_j}X^{(k)}_{\phi_k}\rangle$ (cf. the Supplemental Material).

In conclusion, we have shown how FQH states of impenetrable photons can be generated in a small two-dimensional lattice in a non-equilibrium setting with substantial losses. A particular FQH state is optically selected by the frequency of the coherent pump and the microscopic structure of the wave function can be reconstructed from a few homodyne measurements on the secondary light emission. We hope our results will trigger further work on the intriguing features of FQH physics with photonic systems, including fractional statistics and topological protection of quantum states.

The authors acknowledge financial support from ERC through the QGBE grant.


\begin{thebibliography}{99}
\bibitem{footnote_Dissipation} Alternative strategies to exploit dissipative effects to generate strong correlations in dissipative many-body systems of atoms or photons are discussed, for instance, in~\cite{zoller,syassen} and in~\cite{kiffner}.
\bibitem{footnote_BCs} Note that for the chosen value of $\alpha$ periodic boundary conditions are compatible with invariance under magnetic translations~\cite{Haldane}.
\end{thebibliography}
\end{document}

\title{Supplemental Material}

\maketitle

\renewcommand{\theequation}{S\arabic{equation}}
\setcounter{equation}{0}

\section{Numerical Procedure to Determine the Steady-State of the System}
\subsection{Super-operator Approach}
In this approach, we first transform the time-dependent Hamiltonian $H_0+H_{\rm drive}$ into a time-independent one through a unitary transformation of the form $\mathcal{U} = \exp(i\omega_p t\sum_i \hat{n}_i)$. We then define a super-operator $\mathcal{S}$ in terms of which the master equation can be written as $d\tilde{\rho}/dt = \mathcal{S}\tilde{\rho}$, where $\tilde{\rho}$ is now a vector constructed from the elements of $\rho$. Finding the steady-state value $\tilde{\rho}_{ss}$, which satisfies the equation $d\tilde{\rho}_{ss}/dt = \mathcal{S}\tilde{\rho}_{ss} = 0$, then reduces to finding the eigenvector of the super-operator $\mathcal{S}$ with zero eigenvalue. Note that the size of the matrix representation of $\mathcal{S}$ is $D^2\times D^2$, $D$ being the dimension of the physical Hilbert space, which puts a severe limit on the system size that could be efficiently handled with this method.

\subsection{Monte-Carlo Wave-Function Approach}

In this method (see Ref. [36] of the main text for more details), which can be shown to be equivalent to the master equation treatment, the wave function of the system $|\psi(t)\rangle$ is evolved with the non-Hermitian (time-independent) Hamiltonian $H_0 + H_{\rm drive} - i\hbar\gamma \sum_i\hat{n}_i/2$ and at certain random points in time, depending on the occupation probability of excited states, the system is made to spontaneously emit a photon, or in other words to make a quantum jump. After running this simulation long enough to obtain a sufficient number of statistically independent samples of the density matrix $|\psi(t)\rangle\langle\psi(t)|$, one can get an estimate of the steady-state density matrix by averaging over the samples. The different samples are obtained from a single quantum jump trajectory with sampling intervals on the order of $1/\gamma$ so that the samples can be considered uncorrelated. We checked this method against the super-operator method for several cases, some of which are presented in the main text, and obtained excellent agreement.

\section{Laughlin Wave Function on a Torus}
In an $L\times L$ torus geometry, the $\nu=1/2$ Laughlin wave function of $N$ particles in the Landau gauge $\mathbf{A} = (-By,0)$ has the form
\bea
\label{Laughlin}
\Psi^{(l)}(z_1,...,z_N) &=& \mathcal{N}_L F^{(l)}_{\rm CM}(Z)e^{-\pi\alpha \sum_i y_i^2} \nonumber \\
&\times& \prod_{i<j}^{N}\left(\vartheta \left[
\begin{array}{c}
\frac{1}{2} \\
\frac{1}{2} \\
\end{array}
\right]
\Big(\frac{z_i-z_j}{L}\Big{|}i\Big)\right)^2,
\eea
where $z_k = x_k+i y_k$ is the complex coordinate of the $k$th particle in units of lattice spacing $a$, $Z = \sum_{i}z_i$, $\alpha = Ba^2/(h/e)$ is the number of magnetic flux quanta per plaquette and $\mathcal{N}_L$ is the normalization factor. The part containing relative coordinates is written in terms of the elliptic $\vartheta$ functions $\vartheta [\substack{c\\d}] (z|\tau) = \sum_n e^{i\pi\tau(n+c)^2+2\pi i(n+c)(z+d)}$, where $n$ runs over all integers. The center-of-mass part is also given by elliptic functions:
\be
F^{(l)}_{\rm CM}(Z) = \vartheta \left[
\begin{array}{c}
l/2+(N_{\alpha}-2)/4\\
-(N_{\alpha}-2)/2\\
\end{array}
\right]
\left(\frac{2Z}{L}\Big{|}2i\right).
\ee
Here, $N_{\alpha}$ is the number of flux quanta contained in the $L\times L$ lattice and the label $l =0,1$ indicates the two degenerate ground states at filling $\nu=N/N_{\alpha}=1/2$.

\begin{figure}[h]
\includegraphics[width = \columnwidth]{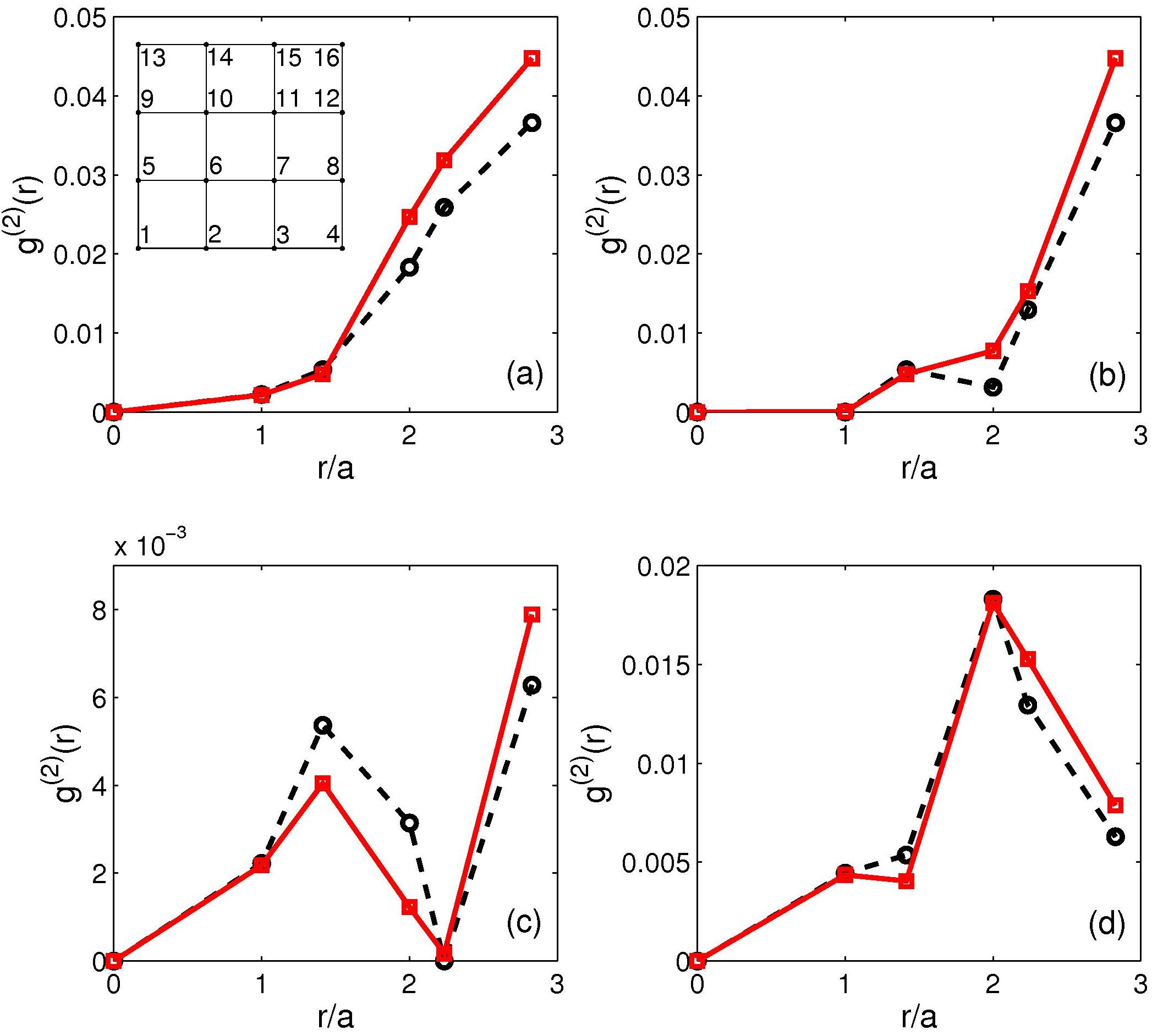}
\caption{(Color online) Second-order correlation function $g^{(2)}(i_\circ,j)$ for different sets of sites $j=\{i_\circ,j_1,\ldots,j_5\}$. Sites are ordered with respect to their distance $r$ from the reference site $i_\circ$. (a) $\{6, 10, 11, 14, 15, 16\}$. (b) $\{6, 7, 11, 8, 12, 16\}$. (c) $\{10, 6, 7, 2, 3, 4\}$. (d) $\{10, 11, 7, 12, 8, 4\}$. Site enumeration is given in the inset of panel (a). Periodic boundary conditions and impenetrable photon limit $U/J=\infty$ are assumed. The pump is on resonance with the two-photon transition to a Laughlin state at $\Delta \omega_p/J=-2.8144$; $\gamma/J=0.01$; $\bar{F}/\gamma=0.1$. Black dashed and red solid lines refer to the Laughlin wave functions and to the numerical results, respectively.
\label{G2}}
\end{figure}

\section{Second-order Correlation Function}

Numerical predictions for the (here unnormalized) second-order correlation function $g^{(2)}(i,j) = \langle\hat{b}^\dag_i\hat{b}^\dag_j\hat{b}_j\hat{b}_i\rangle$ are illustrated in Fig.~\ref{G2} for a $4\times4$ lattice with $\alpha=1/4$ and periodic boundary conditions, and a pump on resonance with the lowest two-photon peak at $\Delta \omega_p/J=-2.8144$. Specifically, we plot $g^{(2)}(i_\circ,j)$ as a function of the distance between site $j$ and a reference site $i_\circ$ for different sets of sites and we compare it with that obtained from the optimized Laughlin wave function which gives the highest overlap. The numerical simulation correctly reproduces the nontrivial behavior of correlations peculiar to FQH states that is visible as peaks and dips in panels (c) and (d).

\begin{figure}[h]
\includegraphics[width = 0.8 \columnwidth,clip]{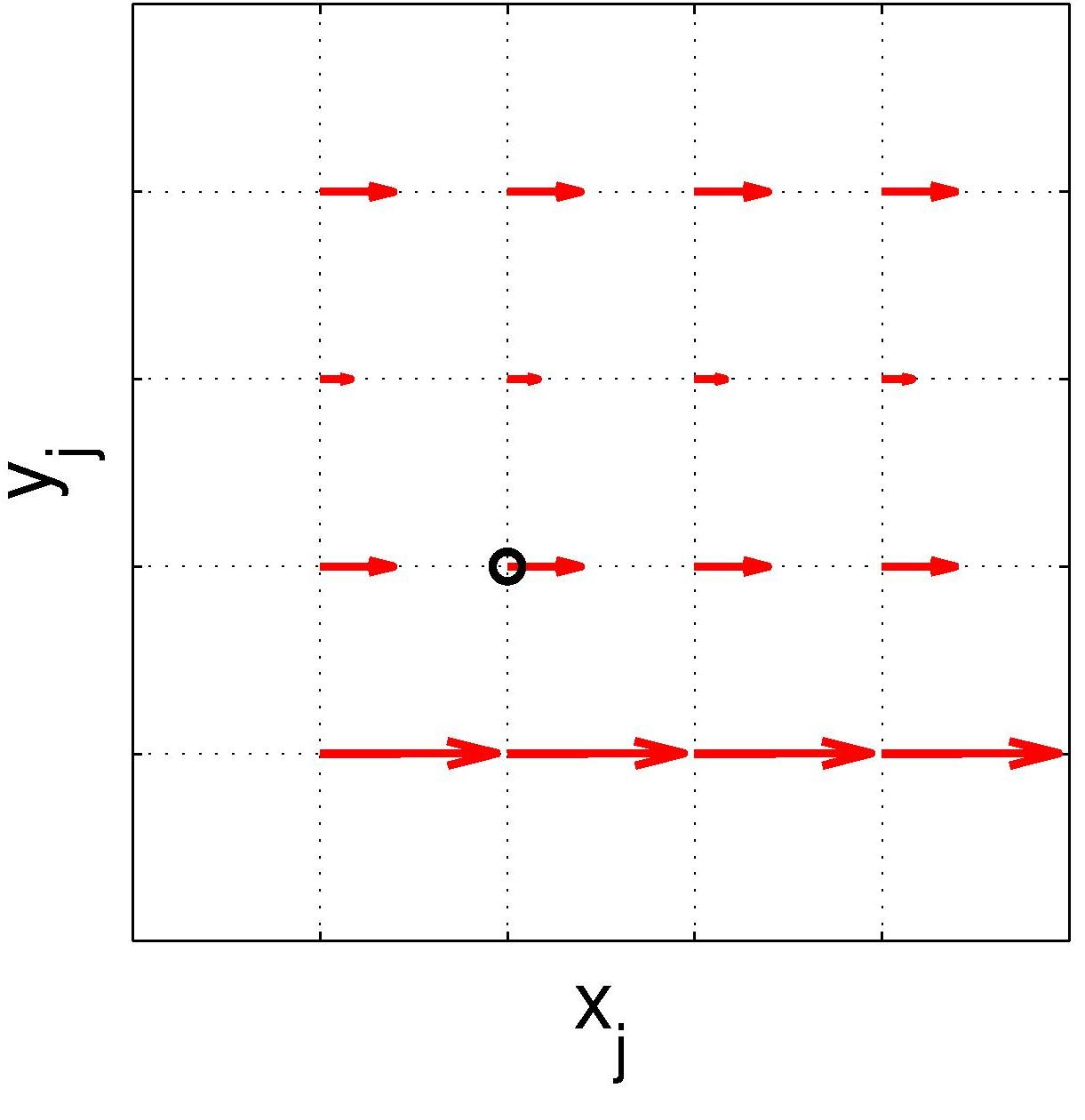}
\caption{(Color online) $i=i_\circ$ cut of the (normalized) two-photon amplitude $\psi_{i_\circ j}$ for non-interacting photons, $U/J=0$. In this case, the cut $\psi_{i_\circ j}$ is proportional to the wave function of the lowest one-particle state $\phi^o_j$ at $\Delta \omega_p/J= -2.8284$. System parameters: $\gamma/J=0.01$; $\bar{F}/\gamma=0.1$. $x,y$ axes refer to the $z_j=x_j+iy_j$ coordinate. Reference site $i_\circ$ is marked by a circle. For each lattice site, the arrow indicates the complex amplitude of the wave function.
\label{4x4correlations}}
\end{figure}

\section{Two-photon Amplitude for Non-interacting Photons}

The two-photon amplitude $\psi_{ij}$ for non-interacting photons with $U/J=0$ is displayed in Fig.~\ref{4x4correlations}. In this case, the two-photon amplitude reduces to the tensor square $\psi_{ij}=\phi^o_i\,\phi^o_j$ of the wave function $\phi_i^o$ of the lowest single-particle eigenstate with zero momentum along $x$ and does not show any special structure in contrast to the Laughlin structure shown in the main text for interacting photons. For the case of a $4\times 4$ lattice with $\alpha=1/4$ under investigation here, each single-particle Hofstadter {\em band} contains a single, four-fold degenerate level. The $k_x=0$ condition imposed by the pump profile selects one single state in the lowest band and non-resonant excitation of states in the upper bands is negligible under the $\gamma/J\ll 1$ condition.

\section{Three-photon Amplitude}

The $N$-photon amplitude $\psi_{i_{1}\ldots i_{N}} = \langle \hat{b}_{i_1}\ldots \hat{b}_{i_N}\rangle$ can be found by using $\hat{b}_i = X^{(i)}_0+iX^{(i)}_{\pi/2}$ obtained from the general definition of the field quadrature $X^{(i)}_{\phi_i} = \big(\hat{b}_i e^{-i\phi_i}+\hat{b}_i^\dag e^{i\phi_i} \big)/2$. So, it is in principle possible to determine this amplitude by doing $2^N$ separate measurements. In particular, the three-photon amplitude can be written as

\bea \label{three_photon} \psi_{ijk} = \langle X^{(i)}_0X^{(j)}_0X^{(k)}_0 \rangle - \langle X^{(i)}_0X^{(j)}_{\pi/2}X^{(k)}_{\pi/2} \rangle \nonumber\\
-\langle X^{(i)}_{\pi/2}X^{(j)}_0X^{(k)}_{\pi/2}\rangle-\langle X^{(i)}_{\pi/2}X^{(j)}_{\pi/2}X^{(k)}_0 \rangle \nonumber \\
-i\langle X^{(i)}_{\pi/2}X^{(j)}_{\pi/2}X^{(k)}_{\pi/2}\rangle+i\langle X^{(i)}_{\pi/2}X^{(j)}_0X^{(k)}_0\rangle \nonumber \\
+i\langle X^{(i)}_0X^{(j)}_{\pi/2}X^{(k)}_0\rangle+i\langle X^{(i)}_0X^{(j)}_0X^{(k)}_{\pi/2}\rangle. \eea